\documentclass[twocolumn,pra,aps,superscriptaddress,showpacs,amsmath,amssymb,floatfix,longbibliography]{revtex4-1}

\usepackage{graphicx}
\usepackage{amsmath}
\usepackage{siunitx} 

\usepackage{bbm}

\usepackage{float}
\usepackage[colorlinks,
linkcolor=red,
anchorcolor=blue,
urlcolor=blue,
citecolor=green]{hyperref}

\begin{document}
	
	
	\title{Universal quantum gates between nitrogen-vacancy centers in a levitated nanodiamond} 
	
	\author{Xing-Yan Chen}
	\affiliation{Center for Quantum Information, Institute for Interdisciplinary Information Sciences, Tsinghua University, Beijing 100084, China}
	\affiliation{Max-Planck-Institut f\"ur Quantenoptik, Garching 85748, Germany}
	
	\author{Zhang-qi Yin}\email{yinzhangqi@tsinghua.edu.cn}
	\affiliation{Center for Quantum Information, Institute for Interdisciplinary Information Sciences, Tsinghua University, Beijing 100084, China}

	
	\date{\today}
	
	\begin{abstract}
	We propose a scheme to realize universal quantum gates between  nitrogen-vacancy (NV) centers in an optically trapped nanodiamond, through uniform magnetic field induced coupling between the NV centers and the torsional mode of the levitated nanodiamond. The gates are tolerant to the thermal noise of the torsional mode. By combining the scheme with dynamical decoupling technology, it is found that the high fidelity quantum gates are possible for the present experimental conditions. The proposed scheme is useful for NV-center-based quantum network and distributed quantum computation.
	\end{abstract}
	
	
	\maketitle 
	
	\section{Introduction}

	Quantum computation is widely believed to be the core of the next generation of information technology. It could be used for solving problems that intractable for classical computers, such as prime factoring \cite{Childs2010}, quantum simulation \cite{Georgescu2014}, machining learning \cite{Biamonte2017}, optimization \cite{Farhi2014}, etc.  In order to realize practical quantum computation, many physical systems have been investigated, such trapped ions \cite{2008PhR...469..155H}, superconducting qubits \cite{PhysRevA.69.062320}, nitrogen-vacancy(NV) centers in diamond \cite{2013PhR...528....1D,Liu2018}, etc. Compared with the other systems, the NV centers can perform quantum tasks, such as individually addressing and initialization, manipulation and detection in room temperature.
	
    In the last decade, there are great progresses in  NV centers based quantum information processing. In NV centers, the electron spins and the nearby nuclear spins couple with each other.  The nuclear spins could be used as long time ($>1$ s) quantum memory at room temperature \cite{Maurer2012}. By using geometric phase and optimizing the controlling pulses, the fault-tolerant universal quantum gates between the electron spins and the nuclear spins have been realized under ambient conditions \cite{Zu2014,Rong2015}. The error correction experiments have been performed in NV center by using both the electron and the nuclear spins, where phase errors have been corrected \cite{Waldherr2014,Taminiau2014}. To correct both phase-flip and bit-flip errors, at least five qubits need to be used, which is experimentally challenge.
    
    The practical quantum computation needs not only fault tolerant, but also scalable. As proposed in Ref. \cite{Jiang2007,Nemoto2014,Monroe2014}, the scalable quantum computation in NV centers could be realized by connecting the small quantum registers with optical channels. 
The NV centers in distant diamonds have been entangled through interference and post selection of the emitted photons \cite{Bernien2013}. Using the similar method, the quantum teleportation \cite{Pfaff2014}, Bell inequality tests \cite{Hensen2015}, entanglement distillation \cite{Kalb2017}, and deterministic entanglement distribution \cite{Humphreys2018} have been realized between the  NV centers in distant diamond. Though  quantum network based on the NV centers develops quickly, the small number of NV centers in every quantum node limits the computational ability of the whole network. 

Up to now, in a single diamond two NV centers electron spins have been entangled through direct coupling \cite{Dolde2013}. 
In order to increase the number of NV centers in a single quantum register, many theoretical schemes have been proposed, such as the dark spin chain data bus \cite{Yao2012,Yao2013}, cavity QED \cite{Park2006,Yang2010}, hybrid NV centers via superconducting circuits \cite{Kubo2010,Marcos2010,Yang2011}, etc. Recent years, there are more and more attentions on the approach of linking NV centers with  mechanical resonators (phonons) \cite{Rabl2010,Strength,Yin2015,Kuzyk2018}. It was found that by using magnetic field \cite{Rabl2009,Yin2013,Ma2017,Delord2017}, or strain effects involving excited states \cite{Golter2016}, the strong coupling between the NV centers and the mechanical phonon modes could be realized. In these systems, the entanglement between multiple NV centers \cite{Xu2009,Zhou2010}, quantum simulation of many-body system \cite{Ma2017}, the single phonon source and detector \cite{Wang2017,Cai2018} and etc.  have been investigated. 
 	
	In this paper, we propose a scheme to realize controlled-phase gate using a levitated nanodiamond where qubits are represented by  NV centers electron spins and the controlled-phase gate is mediated by torsional motion.
The coupling between the electron spins and the torional mode is induced through the uniform external magnetic field \cite{Ma2017,Delord2017}, which is relatively easy to realize compared to the schemes of $10^{7} - 10^{5}$ T/m magnetic gradient induced strong coupling between translational motion and the NV center electron spin \cite{Rabl2009,Yin2013}. Combining with the single qubit operations, we can achieve universal quantum computation in this system. Therefore, as a small universal quantum computer, the levitated nanodiamond with building-in NV centers could be a  quantum node in a large quantum network, which could form a powerful distributed quantum computer.
The two main decoherence effects: rethermalization of the mechanical motion and dephasing of NV centers have been analyzed both analytically and numerically.
	
	\section{The proposed setup}
\begin{figure}[htbp]
		\includegraphics[width=0.8\linewidth]{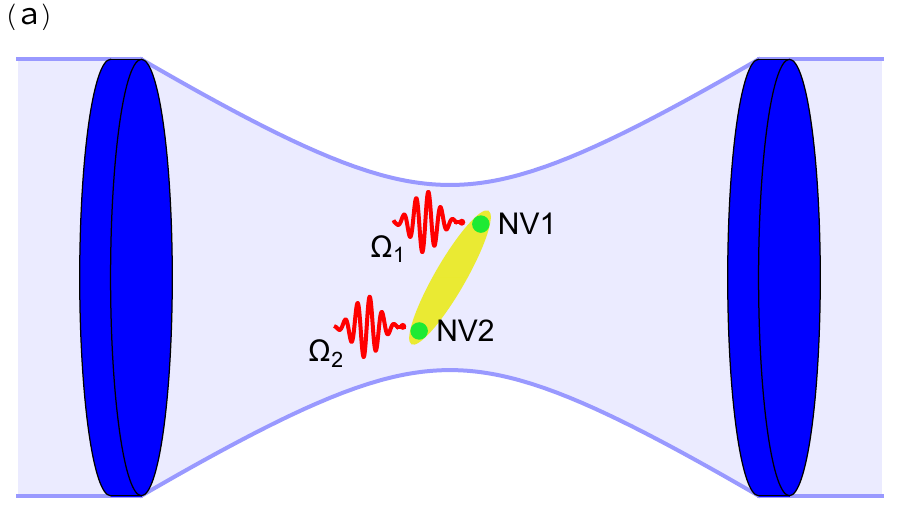}
		\includegraphics[width=0.8\linewidth]{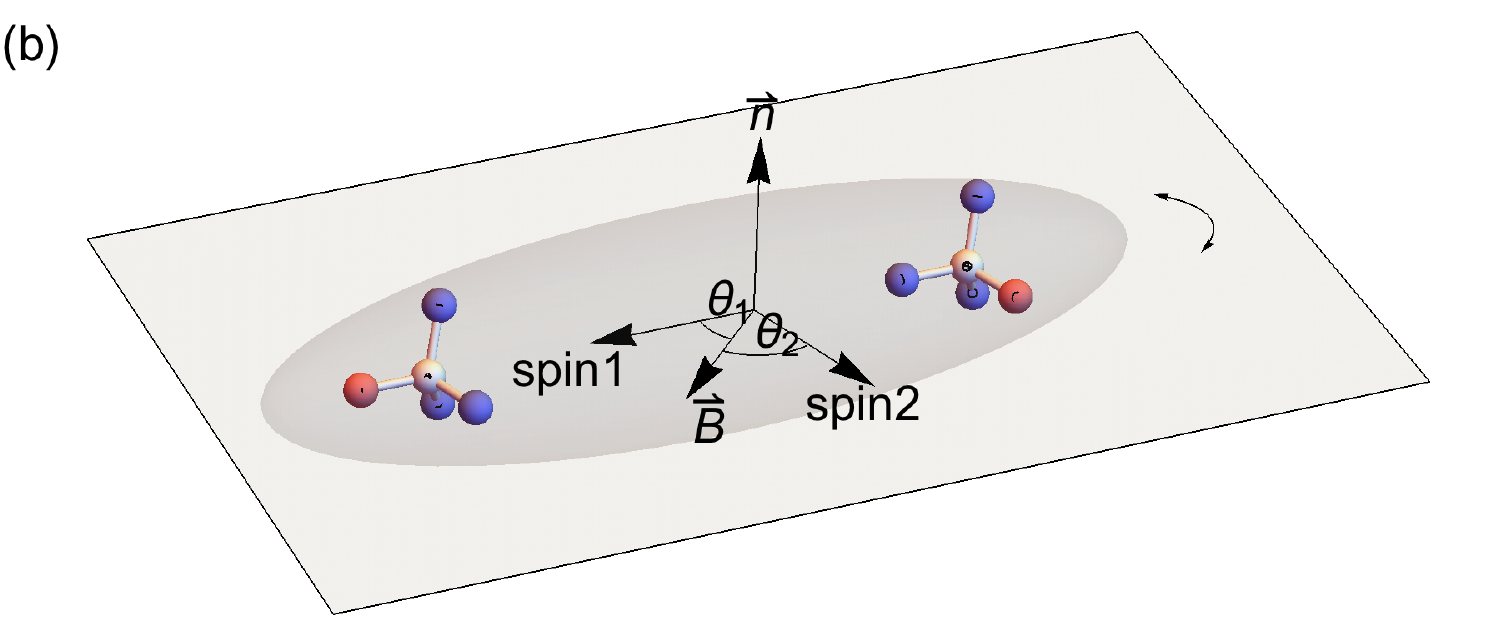}
		\caption{(a) A nanodiamond with two NV centers placed at two endpoints of its long axis is optically trapped in vacuum with spin-torsional coupling through a static magnetic field. (b) Two NV centers in a levitated diamond nanocrystal in the presence of an uniform magnetic field. The rotation orientation $\vec{n}$ is perpendicular to NV centers' spin and the magnetic field $\vec{B}$. Torsional motion of the nanodiamond leads to change in angle $\theta$ which denotes the angle between the orientation of the NV center electron spin$1$ and $\vec{B}$.}		\label{fig:cavity}
	\end{figure}

	We consider a non-spherical nanodiamond with one long axis and two short axes optically trapped in high vacuum in a static uniform magnetic field as shown in Fig.\ref{fig:cavity}(a). Two NV centers are placed at each side of the long axis and along different directions so that their magnetic field induced energy level shifts are different, thus allows single qubit addressing in the frequency domain. Here the qubits are represented by electron spins in the diamond NV centers and the spin-spin coupling is mediated by the librational mode of the nanodiamond. The distance of the two qubits is $d\sim 300$ {nm}  in order to allow individual readout. Besides, as the coupling between the two NV centers scales with $1/d^3$  \cite{Bermudez2011}, in our setup  it is in the order of Hz, much less than the spin-librational-mode coupling. Therefore, we neglect the direct coupling between NV centers in the Hamiltonian of the system.

	We suppose that magnetic field $\vec{B}$ is homogeneous,  and the angle between its direction and the direction of NV center electron spin$1$ $\theta$ changes with the torsional motion. In the energy basis of the NV centers, the total Hamiltonian for the two NV centers and the torsional oscillator reads 
	\begin{equation}
	\begin{split}
	H &= \omega a^{\dagger}a + \frac{E(\theta_1)}{2} \sigma_1^z + \frac{E(\theta_2)}{2} \sigma_2^z +  (g_1 \sigma_1^z + g_2 \sigma_2^z) (a + a^{\dagger}) \\
	&=  \omega a^{\dagger}a + \frac{\omega_1}{2} \sigma_1^z + \frac{\omega_2}{2} \sigma_2^z + \tilde{S} (a + a^{\dagger}),
	\end{split}
	\end{equation}
	where  $\hbar = 1$ (natural unit), $\tilde{S} = g_1 \sigma_1^z + g_2 \sigma_2^z$. $\omega$ is the angular frequency of the torsional mode, $E(\theta_i)$ ($i=1,2$) is the energy splitting between $|s_z = 0\rangle_i \equiv |0\rangle_i $ and $|s_z = -1\rangle_i \equiv |1\rangle_i $ of the spin-1 eigenstates of the two NV centers at relative angle $\theta_i$ to the direction of the magnetic field. The coupling between the torsional mode and the NV center electron spin is
	\begin{equation}
	g_i = \sqrt{\frac{1}{8I\omega_i}}\frac{\partial E(\theta_i)}{\partial\theta_i}, \space \mathrm{i = 1,2},
	\end{equation}
	where $I$ is the moment of inertia of the nanodiamond.
	
\begin{figure}[htbp]
		\centering
		\includegraphics[width=0.9\linewidth]{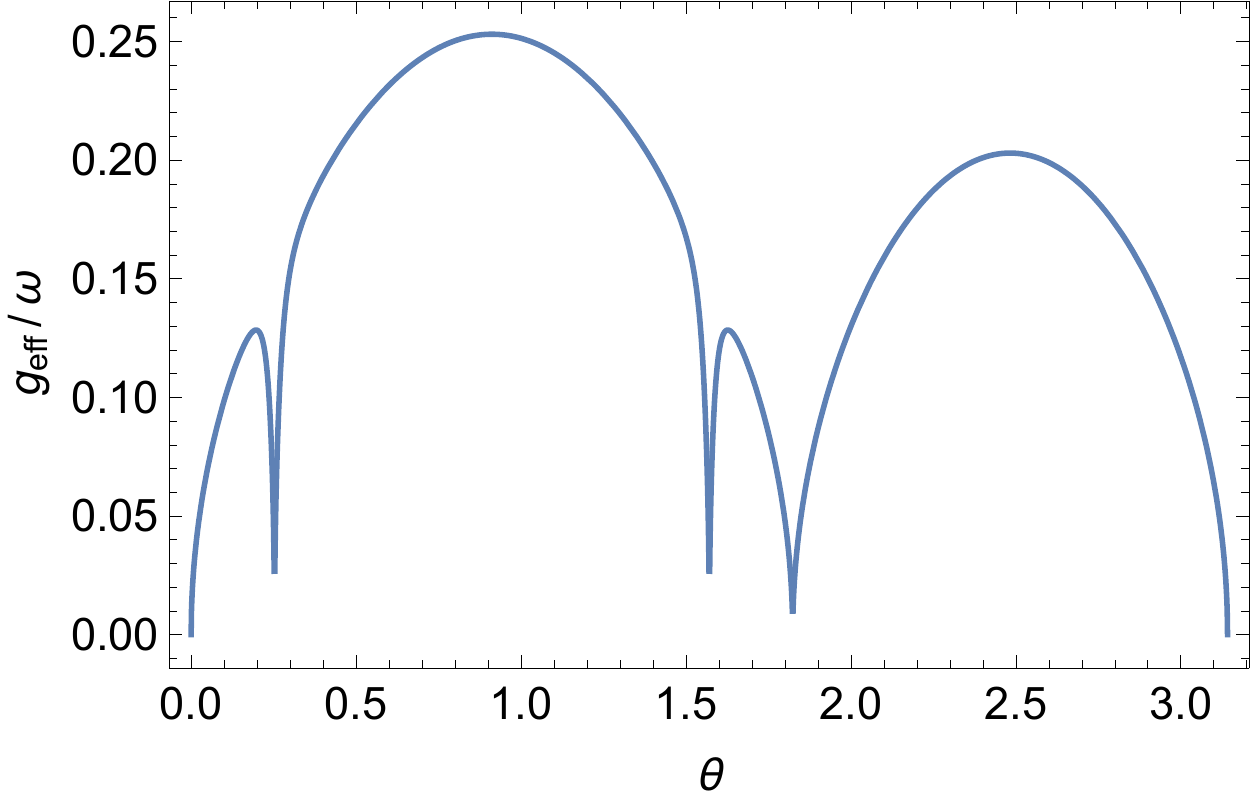}
		\caption{An example of ratio between effective coupling strength $g_{eff} = \sqrt{|g_1g_2|}$ and torsional frequency $\omega$ as a function of relative angle $\theta$, with the magnetic field $B = \SI{7}{mT}$ and $\omega = 2\pi \times \SI{0.1}{MHz}$. The torsional frequency $\omega$ is chosen to fulfill $g/\omega = 1/4$ at the maximum.}
		\label{fig:avgcouple}
	\end{figure}

	In our proposal, the two NV centers are placed along two different directions of the four possible orientations, with an angle $ \theta_1 + \theta_2 = \SI{1.81}{rad}$, as shown in Fig.\ref{fig:cavity}(b). In general, the coupling strengths of the two NV centers $g_1$ and $g_2$  are different. In order to maximize the coupling strength, we control the torsional vibration direction to perpendicular to the two orientations of the NV centers and adjust the magnetic field to be parallel to the vibration plane, as illustrated in Fig.\ref{fig:cavity}(b). The value of $g$ can be tuned in a wide range by controlling either the trap potential or the external magnetic field. We will show in the next section that the gate speed is determined by the effective coupling strength $ g_{eff}^2/\omega = g_1g_2/\omega $. As shown in Fig.\ref{fig:avgcouple}, the averaged coupling strength $ g_{eff} = \sqrt{|g_1g_2|}$ could be about $2\pi \times \SI{25}{kHz}$ at $\SI{0.05}{T}$, which is much larger than  the torsional mode decay ($<$ kHz), the NV center decay ( $<$ kHz), and dephasing rates ($\sim$ kHz). It is comparable to other spin-phonon coupling strength with center of mass motion \cite{Rabl2010,Yin2013,Yin2015}, where a large magnetic gradient $10^5 \sim 10^7 \SI{}{T/m}$ is needed, while our scheme only requires a uniform magnetic field.
	
\begin{figure}[htbp]
		\centering
		\includegraphics[width=\linewidth]{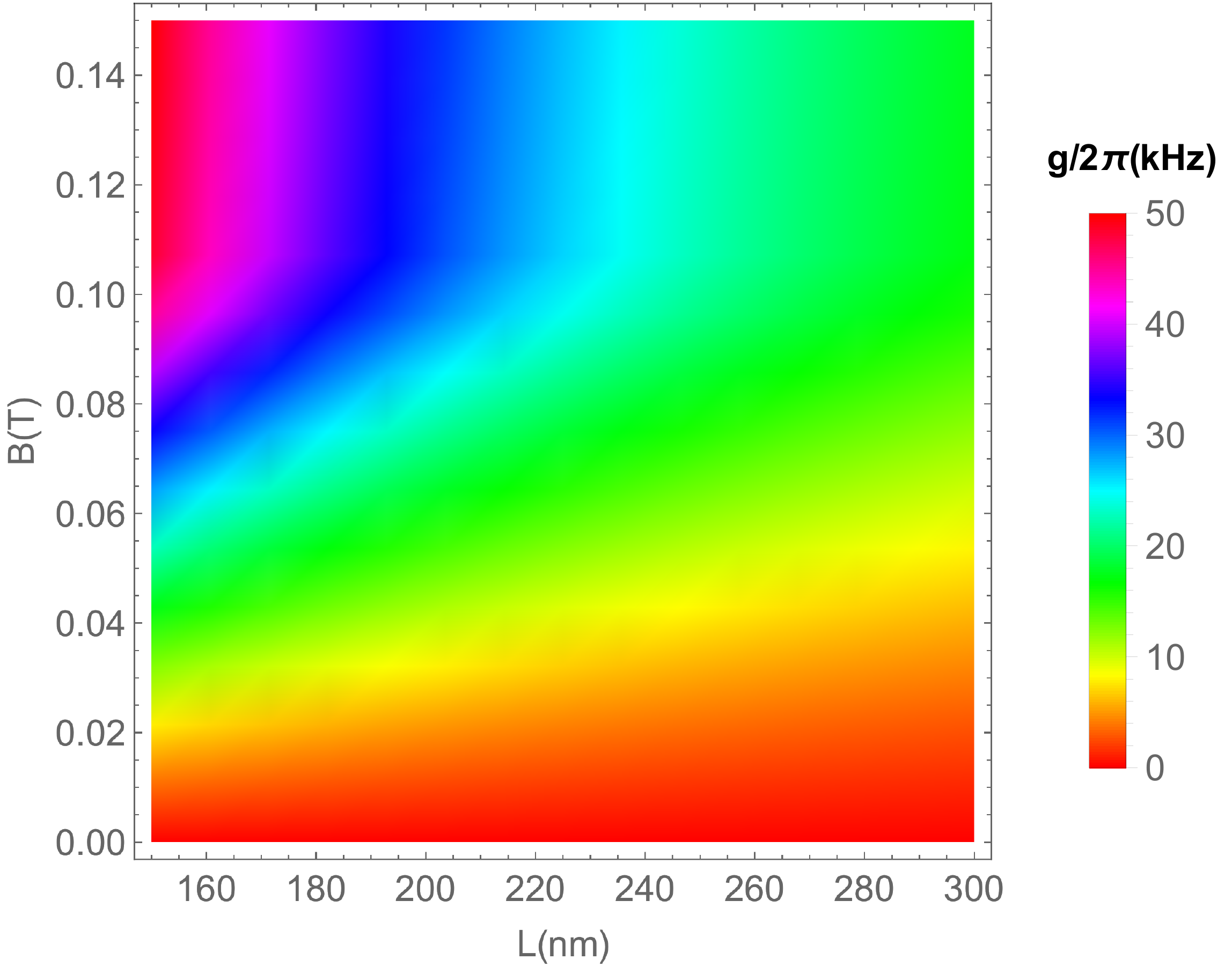}
		\caption{The coupling strength $g$ under different length $L$ of the nanodiamond and the external magnetic field $B$. The ratio of the long and short axis is fixed to be $1.5$. The torsional frequency is taken to be $2\pi \times \SI{.1}{MHz}$ at $L = \SI{150}{nm}$. The coupling strength is optimized along all the relative direction of the spin to the magnetic field, e.g. take the maximum value in Fig.\ref{fig:avgcouple}.}
		\label{fig:size}
	\end{figure}

	In principle, the nanodiamond can contain multiple NV centers which couple to  the same torsional mode. The NV centers should be separated by more than $\SI{100}{nm}$ to suppress unwanted dipole-dipole interactions and support the individual readout \cite{zhang2017selective}. Therefore in order to embedding more NV centers, the size of the nanodiamond should increase, which would decrease the torsional frequency and reduce the coupling strength. The size effect on the coupling strength is shown in Fig.\ref{fig:size}. When the long and short axis are $\SI{300}{nm}$ and $\SI{200}{nm}$, respectively, the number of the embedded NV can be about $10$, and the electron spins torsional mode coupling $g$ around $2\pi \times 20$ kHz.
	
	\section{Gate operation under ideal conditions}
	Single qubit operations can be realized by simply adding microwave pulses  to induce transition between the two energy levels which represent $|0\rangle$ and $|1\rangle$ with proper phase. If the Rabi frequency of the drive is large enough, we can ignore the term $g \sigma_z (a + a^{\dagger})$ which brings about unwanted coupling to the vibration. Another way to decouple the qubit states from the resonator is to incorporate dynamical decoupling \cite{Khodjasteh2010,Cohen2017} or geometric approach \cite{Sjoqvist2012,Zu2014} which simultaneously suppress the noise caused by a nuclear-spin bath. Further description and numerical simulation on noise effect will be presented in the section \ref{sec:Dec}.
	
	Here we propose a method to realize controlled phase gates and universal computation could be achieved when combined with single qubit operations. Our method is inspired by  the celebrated S{\o}rensen-M{\o}lmer gates for hot trapped ions \cite{Molmer1999,Sorensen2000} or similar bus-based quantum gates \cite{Kolkowitz2012b,Rabl2010,Yin2013,Schuetz2017}.
		In the interaction picture of $ A = \frac{\omega_1}{2} \sigma_1^z + \frac{\omega_2}{2} \sigma_2^z$, we obtain the interaction Hamiltonian 
\begin{equation}	\label{eq:HI}
  \begin{split}
	H' &= \omega a^{\dagger}a + \tilde{S} (a + a^{\dagger}) \\
	&= \omega (a + \frac{\tilde{S}}{\omega})^{\dagger}(a + \frac{\tilde{S}}{\omega}) - \frac{\tilde{S}^2}{\omega} 
  \end{split}
\end{equation}
	Using similar method as \cite{Schuetz2017}, the Hamiltonian \eqref{eq:HI} can be rewritten as
	\begin{equation}
	H' = U [\omega a^\dagger a - \frac{\tilde{S}^2}{\omega}] U^\dagger,
	\end{equation}
	where $U$ is unitary transformation $U = \exp[\frac{\tilde{S}}{\omega}(a-a^\dagger)]$. 
	The time evolution governed by the Hamiltonian \eqref{eq:HI} reads 
	\begin{equation}
	e^{-iH't} = U e^{-i\omega t a^\dagger a}e^{i\frac{\tilde{S}^2}{\omega}t}U^\dagger.
	\end{equation}
	For certain times when $ \omega t_m = 2\pi m $ with $m=1,2,3\cdots$, the first exponential 
$e^{-i\omega t_m a^\dagger a} = \mathbbm{1} $ since the number operator has an integer spectrum. Thus time evolution reduces to
	\begin{equation}
	e^{-iH't_m} = \exp \big[  2i \pi m (\frac{\tilde{S}}{\omega})^2 \big].
	\end{equation}
	Substituting $ \tilde{S}^2 = g_1^2 + g_2^2 + 2g_1g_2 \sigma_1^z\otimes\sigma_2^z  $, and ignoring the unimportant global phase we get
	\begin{equation} \label{eq.cphase}
	e^{-iH't_m} = \exp( 4i\pi m \frac{g_1g_2}{\omega^2} \sigma_1^z\otimes\sigma_2^z).
	\end{equation}
	
Returning to the laboratory frame, the full evolution is governed by $e^{-iH't_m}e^{-iAt_m}$. We now show that the $e^{-iAt_m}$ term can be canceled exactly by the standard spin-echo technique, which compensates the unknown detuning as well. Denoting a global flip of all qubits around the axis $\alpha = x,y,z$ as $U_\alpha(\varphi) = exp(-i\varphi/2\Sigma_i \sigma_i^\alpha)$, the full evolution (in the basis $\{|00\rangle,|10\rangle,|01\rangle,|11\rangle\}$), intertwined by spin-echo pulses, reads as
	\begin{align}
	U(2t_m) &= U_x(\pi)e^{-iH't_m}e^{-iAt_m}U_x(\pi)e^{-iH't_m}e^{-iAt_m} \\
	&= \text{diag}(e^{i\phi},1,1,e^{i\phi}),
	\end{align}
	with $\phi = 8m\pi(g_{eff}/\omega)^2$ and $g_{eff} = \sqrt{|g_1g_2|}$. After complementing the propagator $U(2t_m)$ with $U_z(-\phi)$, we obtain
	\begin{align}
	U_{\mathrm{Cphase}} &= U_z(-\phi)U(2t_m) \\
	&= \text{diag}(1,1,1,e^{2i\phi})
	\end{align}
	which is a controlled phase gate for $\phi = \pi/2$, corresponding to a gate time $t_{max} = 2t_m = 4\pi m/\omega = \pi/4g_{eff}$. Here $m$ must take integer value. The condition $\phi = \pi/2$ can be archieved when 
	\begin{equation}\label{eq.cond}
	g_{eff}/\omega = \frac{1}{4\sqrt{m}}.
	\end{equation} which gives $g_{eff}/\omega = 1/4$ for $m = 1$.
	To fulfill the above condition we can change $ \omega $ by tuning the power and waist of the optical tweezer. As an example, in Fig. \ref{fig:avgcouple}, for given magnetic field, we can choose appropriate $\omega$ to approximate condition Eq.(\ref{eq.cond}) with $m=1$ in a wide range (from $0.8$ to $1.0$ rad) of direction $\theta$.

	\section{Decoherence effect} \label{sec:Dec}
	
	The main sources of decoherence in this hybrid system of solid-state spins and mechanical oscillator are the rethermalization of the torsional mode towards equilibrium thermal occupation and the dephasing effect of the spins. The dissipative dynamics under these decoherence effects can be described by the master equation for the system's density matrix $\rho$,
	\begin{align}\label{eq.master}
	\dot{\rho} =& -i [H,\rho] + \kappa(\bar{n}_{th} + 1)\mathcal{D}[a]\rho + \kappa\bar{n}_{th}\mathcal{D}[a^\dagger]\rho \nonumber \\
	& + \frac{\Gamma}{4}\sum_{i=1,2}\mathcal{D}[\sigma_i^z]\rho
	\end{align}
	with $\mathcal{D}[a]\rho = a\rho a^\dagger - \frac{1}{2}\{a^\dagger a,\rho\}$ and the torsional mode decay rate $\kappa = \omega/Q$, where $Q$ is the quality factor of the torsional mode. The second and the third terms in Eq. \eqref{eq.master} describe the rethermalization of the torsional mode towards the thermal occupation $\bar{n}_{th}=[\exp(\hbar\omega/k_BT)-1]^{-1}$ at temperature $T$. The last term in Eq. (\ref{eq.master}) describes the dephasing of the qubits with a dephasing rate $\Gamma \sim 1/T_2 $, where $T_2 $ is the time-averaged dephasing time. Here we have ignored single-spin relaxation processes as the associated time scale $T_1$ is typically much longer than $T_2$. Nevertheless, by using dynamically  decoupling techniques \cite{Cai2012,Rong2014,Cohen2017,Cohen2017a} the coherence time can be efficiently prolong to approach $T_1$.
	
	In the case of single qubit gates, the coupling between the mechanical mode and the spin $g\sigma_i^z(a+a^\dagger)$ serves as an extra dephasing channel. The coupling induced dephasing effect, characterized by $\delta_g$, grows with the thermal occupation $\bar{n}_{th}$, and is much larger than the effect of nuclear-spin bath $\delta_S$ under typical experimental parameters $\bar{n}_{th} \sim 1\times 10^3$. Therefore, in the following section we introduce a dynamical corrected operation which can significantly reduce both dephasing effects.
	\subsection{single qubit gates}
	
	\begin{figure}[htbp]
		\centering
		\includegraphics[width=0.9\linewidth]{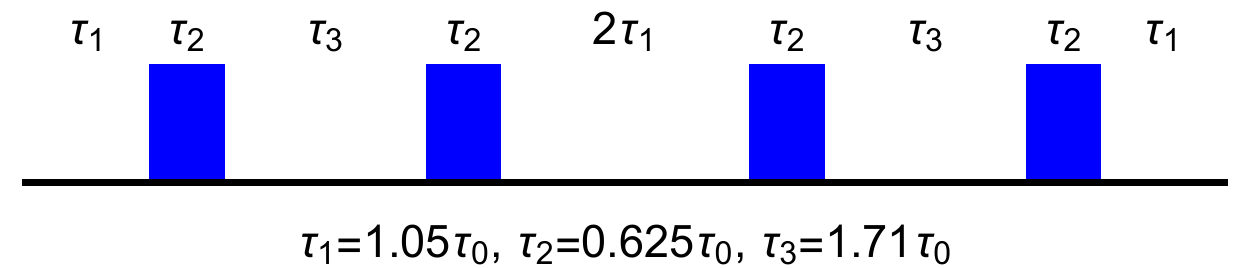}
		\caption{Pulse sequences of 5-piece SUPCODE sequences \cite{Rong2014}. The time duration $\tau_0 = (2+\theta/2\pi)t_0$ for a $\theta$ rotation pulse.}
		\label{fig:supcode}
	\end{figure}	
	For single qubit operations, we consider an on-resonance drive along $x$ axis with Rabi frequency $(\Omega \sim 2\pi\times 50$ MHz. The Hamiltonian can be written as $H=2\pi(\delta S_z+\Omega S_x)$, with $\delta = \delta_S + \delta_g$ represents the effect of interaction with the nuclear-spin bath and coupling to the mechanical motion. The hyperfine interaction with the nuclear spin results in a random local magnetic field $\delta_S = \Sigma_k b_k I_z^k $ of typical strength of the order of magnitude of about $1$ MHz in solids. The coupling effect of the mechanical mode can be estimated by $ \delta_g = g(a + a^\dagger) \sim 2g\sqrt{\bar{n}_{th}} $ for $\bar{n}_{th} \gg 1$. With a five-piece SUPCODE pulse described in \cite{Rong2014}, the infidelity of the $\pi$ gate can be estimated by $ 64.1(\delta/\Omega)^6 + O(\delta/\Omega)^8 $. The 5-piece pulse sequence is shown in Fig.\ref{fig:supcode}.

	\begin{figure}[htpb]
		\includegraphics[width=0.85\linewidth]{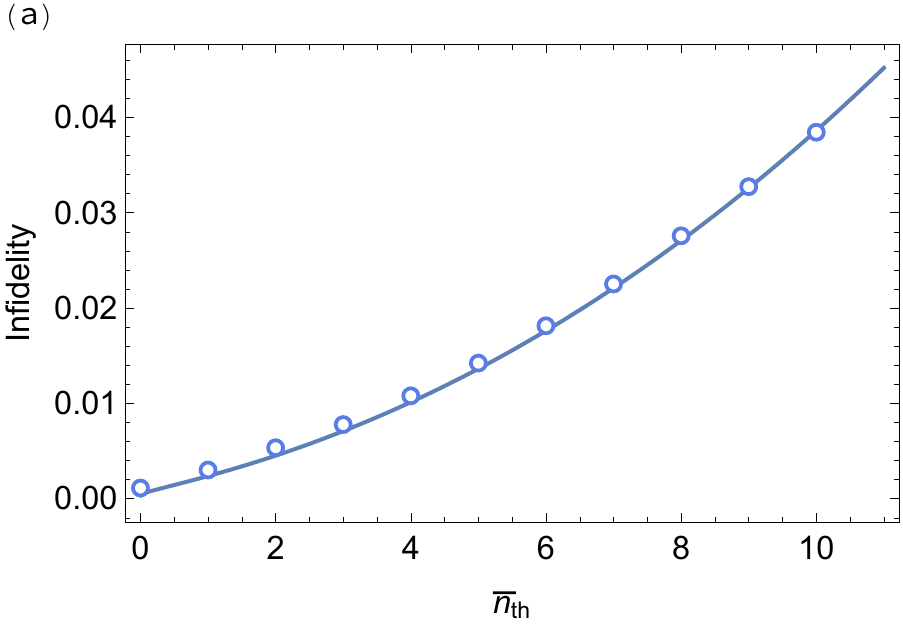}
		\includegraphics[width=0.85\linewidth]{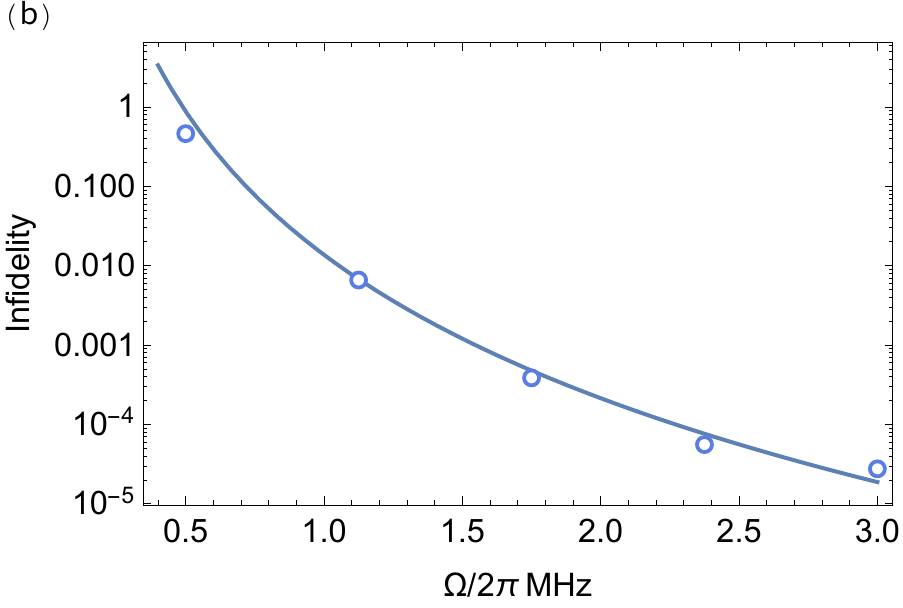}
		\caption{Infidelity due to coupling to torsional mode as a function of (a) thermal occupation $\bar{n}_{th}$ with Rabi frequency $\Omega = 2\pi\times \SI{1}{MHz}$; (b) Rabi frequency $\Omega$, with thermal occupation $\bar{n}_{th}=5$. The torsional frequency $\omega = 2\pi\times \SI{0.1}{MHz}$ and coupling strength $g = \omega/4$. Blue circles are numerical results obtained by solving the master equationn, and the blue line is the fitted model of Eq.(\ref{eq.err_delta}) with $n_0 = 2.65$.}
		\label{fig:singlenth}
	\end{figure}
		
	In the following, we provide the numerical results of the master equation Eq.(\ref{eq.master}), for the five-piece SUPCODE $\pi$ pulse around x(y) axis. For thermal occupation $\bar{n}_{th} = 1000$, the dephasing induced by coupling to the resonator mode $\delta_g\sim {4} \SI{}{MHz}$ while the typical strength of the random magnetic field $\delta_S \sim 1$ MHz. We focus on the coupling effect here, and leave the discussion of rethermalization and dephasing in the next section. Typical results from our numerical simulations are displayed in Fig.\ref{fig:singlenth}. The results can be well explained by a fit to
	\begin{equation}\label{eq.err_delta}
	\xi = 1 - \mathcal{F} =64.1(\frac{2g(\sqrt{\bar{n}_{th}}+n_0)}{\Omega})^6,
	\end{equation}
	which is obtained by replace $ g(a + a^\dagger)$ by $2g(\sqrt{\bar{n}_{th}}+n_0)$ in the noise parameter $\delta$. The parameter $n_0$ represents the thermal fluctuation which still exists at thermal vacuum.

	By Eq.(\ref{eq.err_delta}), even with large thermal occupation $\bar{n}_{th} = 300$, infidelity $\sim 10^{-3}$ can be achieved with Rabi frequency $ \Omega \sim 2\pi\times \SI{10}{MHz}$.
	
	\subsection{controlled phase gate}	
\begin{figure}[htbp]
		\includegraphics[width=0.85\linewidth]{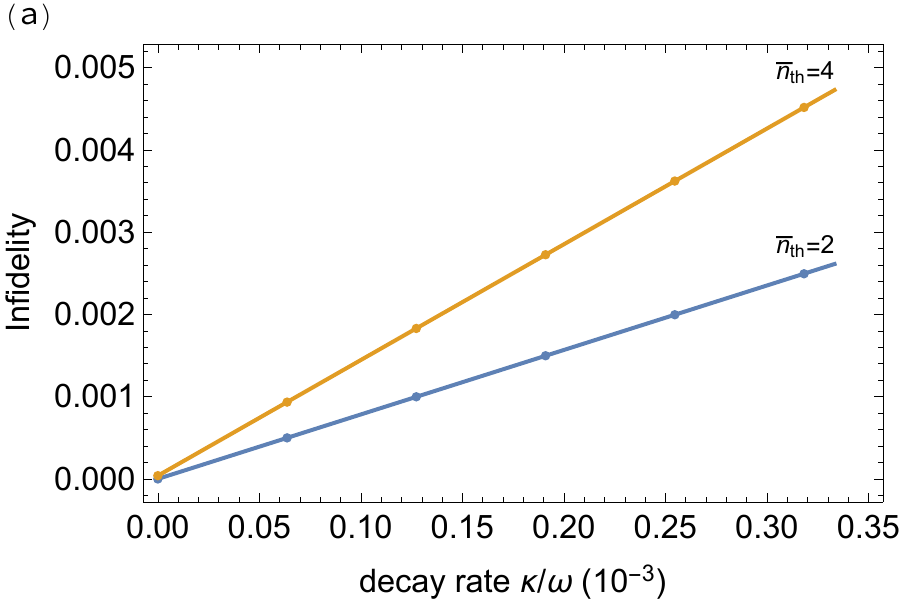}
		\includegraphics[width=0.85\linewidth]{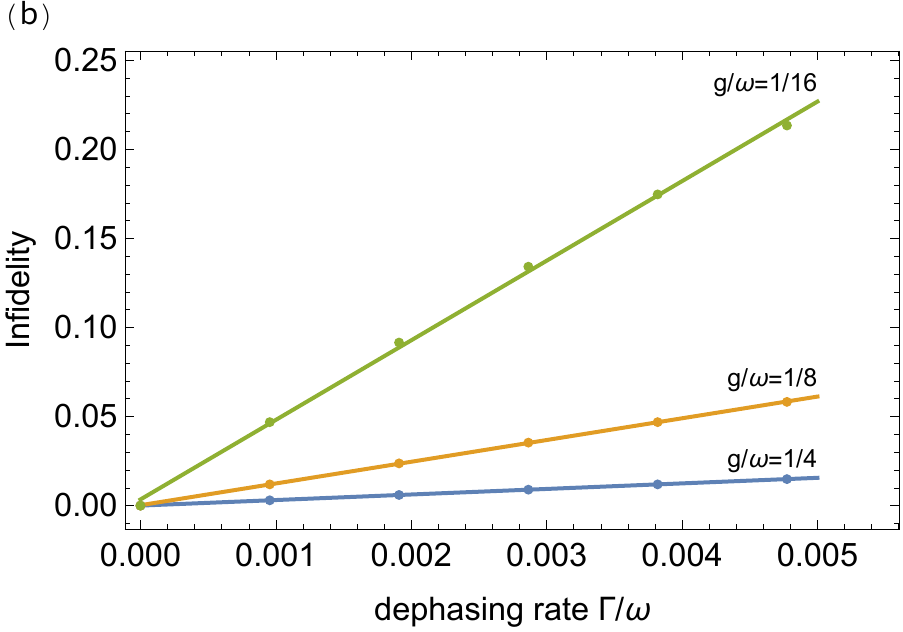}
		\caption{Errors ($\xi = 1 - \mathcal{F}_{max}$) due to rethermalization of the cavity mode (a) and qubit dephasing (b). (a) Rethermalization-induced error for $ \bar{n}_{th} = 2$ (blue) and $\bar{n}_{th} = 4$ (orange), and $\Gamma = 0$.
			(b) Dephasing-induced errors for $\mu = g/\omega = 1/4$ (blue), $\mu = 1/8 $(orange), and $\mu = 1/16$ (green); here, $\kappa/\omega = 10^{-6}$ and $ \bar{n}_{th} = 0.01$. In both cases, the linear error scaling is verified.}
		\label{fig:retherm}
	\end{figure}

	For controlled phase gate, we provide numerical results of the master equation (\ref{eq.master}), for the initial product state $\rho(0) = (|00\rangle + |11\rangle)(\langle00| + \langle11|)\otimes\rho_{th}(T) $. We quantify the state fidelity $\mathcal{F}=\langle\Phi_{tar}|\varrho|\Phi_{tar}\rangle$ with the target state $|\Phi_{tar}\rangle = (|00\rangle - |11\rangle)/\sqrt{2}$; here, $\varrho=tr_a[\rho]$ refers to the density matrix of the qubits, with $tr_a[...]$ denoting the trace over the resonator degrees of freedom. Here we use dynamical decoupling pulse described in \cite{West2010,Ng2011} which can suppress the decoherence rate $\Gamma$ by two or three orders of magnitude as compared with the pure dephasing rate.
	In order to study the individual effects of the decoherence channel, we treat them separately here, and assume perfect single qubit operation. This separate treatment is justified by comparing the sum of individual errors with the results from full master equation. Typical results from our numerical simulations are displayed in Fig.\ref{fig:retherm}. For small infidelities $(g_\text{eff} \gg k_\text{eff},\Gamma)$, the individual linear error terms due to cavity rethermalization and qubit dephasing can be added independently. Using the simple linear error model in \cite{Schuetz2017}, the total error reads
	\begin{equation}
	\xi \approx \alpha_\kappa/Q \bar{n}_{th} + \alpha_\Gamma \Gamma/\omega.
	\end{equation}
	Based on results in Fig. \ref{fig:retherm} we extract the coefficient $\alpha_\kappa \approx 4 $ and $\alpha_\Gamma \approx 0.2(\omega/g_{eff})^2 $. Here the coefficient related to dephasing  $\alpha_\Gamma $ is different from the result in \cite{Schuetz2017} by a factor of $2$, as the controlled-phase gate time in our proposal is twice as the time as it is to obtain the maximally entangled state. We can estimate an upper limit of the thermal occupation number $n_th$ with typical parameters in our system are $ \omega = 2\pi \times \SI{1}{MHz} $, $ g/\omega = 1/4 $. For a high Q torsional mode with $Q = 10^{11}$, the rethermalization effect can be ignored even at room temperature(corresponding to $\bar{n}_{th}\approx 8\times 10^6$) . Nevertheless, in order to avoid an-harmonic effect of large torsional amplitude $\Delta \theta$, the condition $ \frac{1}{2}\partial^2 E/\partial \theta^2 (\Delta \theta)^2 \ll \partial E/\partial \theta \Delta \theta $ has to be fulfilled which corresponding to $ \bar{n}_{th} \ll 3\times 10^6$. 
In addition to rethermalization, the effective dephasing rate given a pure dephasing rate 100kHz is ~1kHz \cite{Cohen2017,Cohen2017a}, which gives the estimated error $\xi \approx 0.3\%$.
	
	\section{Conclusions}
	We have proposed a scheme to realize a high-fidelity universl quantum gates in a spin-torsional motion system, even in the presence of a thermally populated torsional mode. Our proposal uses a uniform magnetic field instead of a large magnetic gradient and thus reduce the noise caused by fluctuation of magnetic field. The spin-torsional coupling can be larger than $\SI{100}{kHz}$ which allows a short gate time less than $\SI{8}{\mu}$s for a controlled-phase gate. The two main decoherence effects: rethermalization of the mechanical motion and dephasing of NVs have been analyzed both analytically and numerically. The mechanical motion serves as the quantum bus for two qubit controlled-phase gates while induces extra dephasing channel for single qubit gates. To increase the coherence time, we have incorporated dynamical decoupling method for single qubit gates. It is found that the high fiedlity larger than $99\%$ universal quantum gates are possible under the current experimental conditions. Our scheme could be applied for distributed quantum computation and quantum network based on NV centers. 

\begin{acknowledgments}
This work is supported by National Natural Science Foundation of China NO. 61771278, 61435007, and the Joint
Foundation of Ministry of Education of China (6141A02011604).
We thank the helpful discussions with  Tongcang Li.

\end{acknowledgments}
	
%
	
\end{document}